\documentclass[smallabstract,smallcaptions]{dccpaper}
\usepackage[T1]{fontenc}
\usepackage{graphicx}
\usepackage{hyperref}       
\usepackage{amsfonts}       
\usepackage{amsmath}		
\usepackage{cleveref}       
\usepackage[ruled,lined,linesnumbered, noend]{algorithm2e}
\usepackage{nowidow}
\DontPrintSemicolon{}
\newcommand{\bigO}{\mathcal{O}}
\newcommand{\state}[1]{\(#1\) \;}
\newcommand{\assign}[2]{\state{#1 \es{} \leftarrow{} \es{} #2}}
\newcommand{\es}{\textbf{ }}
\newcommand{\firstTerminalName}{g}
\newcommand{\secondTerminalName}{f}
\newcommand{\firstNonterminalName}{B}
\newcommand{\secondNonterminalName}{C}

\usepackage{color}
\usepackage{enumitem}
\setlist{nolistsep}

\makeatletter
\def\moverlay{\mathpalette\mov@rlay}
\def\mov@rlay#1#2{\leavevmode\vtop{%
   \baselineskip\z@skip \lineskiplimit-\maxdimen
   \ialign{\hfil$\m@th#1##$\hfil\cr#2\crcr}}}
\newcommand{\charfusion}[3][\mathord]{
    #1{\ifx#1\mathop\vphantom{#2}\fi
        \mathpalette\mov@rlay{#2\cr#3}
      }
    \ifx#1\mathop\expandafter\displaylimits\fi}
\makeatother

\newcommand{\bigcupdot}{\charfusion[\mathop]{\bigcup}{\cdot}}

\begin{document}
\title{\large \textbf{ITR:\@ Grammar-based Graph Compression Supporting Fast Triple Queries}}
%
%
%
%
%
\author{Enno Adler,
Stefan Böttcher,
Rita Hartel\\[0.5em]
    {
    \small\begin{minipage}{\linewidth}
        \begin{center}
            Paderborn University,\\
            Paderborn, Germany\\
            {\{enno.adler,stefan.boettcher,rita.hartel\}@uni-paderborn.de}\\[0.5em]
        \end{center}
    \end{minipage}
    }
}
\maketitle              

\begin{abstract} 
    Neighborhood queries and triple queries are the most common queries on graphs; thus, it is desirable to answer them efficiently on compressed data structures. We present a compression scheme called Incidence-Type-RePair (ITR) for graphs with labeled nodes and labeled edges based on RePair~\cite{larsson2000} and apply the scheme to network, version, and RDF graphs. We show that ITR performs neighborhood queries and triple queries within only a few milliseconds and thereby outperforms existing RePair-based solutions on graphs while providing a compression size comparable to existing graph compressors.
\end{abstract}

\section[Introduction]{Introduction}
Obtaining a node or all nodes in a graph that are adjacent to a given node is fundamental to most graph algorithms. Therefore, these neighborhood queries are the most common queries in graph processing. Whenever huge graphs, for example, network, version, or RDF graphs, are compressed, and neighborhood queries are heavily used on these compressed graphs, their performance is crucial for improving the efficiency of graph processing and the analysis of large-scale graphs. We investigate the execution time of neighborhood queries and the more general triple queries, which are the base of graph databases, on compressed graphs generated by different graph compressors. Furthermore, we introduce Incidence-Type-RePair (ITR), a grammar-based compressor that generates a compressed graph on which nearly all triple queries are answered significantly faster than on other compressed data formats for these graphs.

Like RePair~\cite{larsson2000}, ITR uses context-free grammars for compression and repeatedly replaces a most frequent digram by a new nonterminal. The term \textit{digram} describes two adjacent elements. For a string $ababab$, $ab$ is a digram of two adjacent letters and the grammar $\{S \rightarrow AAA, A \rightarrow ab\}$ results from replacing $ab$ by $A$ in $ababab$.

Grammar-based compression schemes have been shown to improve the efficiency of queries on compressed data, for example, for consecutive symbol visits on strings~\cite{gasieniec2005} and for parent/child navigations on trees~\cite{Lohrey2011, lohrey2016}. 
Furthermore, Maneth et al.~\cite{Maneth2016} and Röder et al.~\cite{roeder2021} both apply RePair to graphs and define digrams to be two edges sharing a common node. As explained in Maneth et al.~\cite{Maneth2016}, finding a largest possible set of non-overlapping occurrences for a single digram, requires $\bigO{(|V|^2 |E|)}$ time and is thus infeasible. Instead, they and we present different approximations on how to define and find frequent digrams.

\newpage{}

The main contributions of this paper are: 

\begin{itemize}
    \item ITR, an algorithm combining labeled edges and node labels to hyperedges;
    \item a digram definition that simplifies and speeds-up finding and replacing all occurrences of a most frequent digram;
    \item the index-functions, which optimize the compression of loops within edges;
    \item the substitution of frequent node labels by hyperedges to reduce the number of dictionary entries;
    \item an evaluation comparing ITR to gRePair, RDF\-RePair, and different implementations of HDT~\cite{fernandez2013} and $k^2$-tree~\cite{Brisaboa2009} regarding compression size and runtime of neighborhood queries and triple queries that shows that ITR for nearly all types of triple queries outperforms the other implementations of graph compressors.
\end{itemize}

\section[Preliminaries]{Preliminaries}

A \textit{ranked alphabet} $\Sigma$ is a set of symbols with a function $rank:\Sigma \rightarrow \mathbb{N}\backslash\{0\}$ that maps each symbol to its rank. Let $L$ be a ranked alphabet called labels. A \textit{hypergraph} is a pair $G = (V_G, E_G)$ with nodes $V_G = \{0, 1, \ldots, p\} \subset \mathbb{N}$ and edges $E_G\subset L \times V_G^*$.\footnote{$V_G^*$ is the set of all possible sequences of nodes from $V_G$.} We write $e = a(v_0, \ldots, v_{n-1})$ for an edge $e\in E_G$ with $label(e)=a$ and $rank(e) = n$. The rank describes the number of nodes (including duplicates) that are connected to the edge. We call all elements in $E_G$ edges regardless of their rank, and we use the term hyperedge to emphasize that a concept requires non-rank-2 edges. Let $\mathbb{G}$ be the set of all hypergraphs. 

We write $e[m] = v_m$ for $v_m$ that is connected to $e=a(v_0, \ldots, v_{n-1}) \in E_G$, and we call $m$ the \textit{connection-type} of $v_m$ to the edge $e$. We always assume $m$ to be well defined, that is $0\leq m < n = rank(e)$. For example, an edge $e=\secondTerminalName(10, 11) \in E_G$ has $label(e)=\secondTerminalName$, $rank(e)=2$, and connection-type $0$ for node $e[0] = 10$. For each edge of rank 2, there are two connection-types, where 0 is equivalent to `outgoing' from the source node and 1 is equivalent to `incoming' to the destination node. For all symbols $a \in L$ and for all edges $e\in E$ with $label(e)=a$, we require $rank(e) = rank(a)$, that is, all edges with the same label have the same rank. These unique ranks for edges with the same label are necessary to define the rules well.

Similar to Maneth et al.~\cite{Maneth2016}, we define a \textit{hyperedge replacement grammar} (HR grammar) as a tuple $H = (T, N, P, S)$ with $T$ and $N$ being ranked alphabets with $N \cap T = \emptyset$, $T \cup N = L$, $P \subset N\times \mathbb{G}$, and $S \in N$. We call an edge $e$ terminal if $label(e) \in T$, and we call $e$ nonterminal if $label(e) \in N$. We write $A \rightarrow G_A$ for a rule $(A, G_A) \in P$. The right-hand side $G_S$ of the grammar rule $S \rightarrow G_S$ is called \textit{start graph}.

We construct only \textit{straight-line} hyperedge replacement grammars (SL-HR grammars). For straight-line grammars, the following conditions hold: (1) for each nonterminal in $N$ there exists exactly one rule in $P$ and (2) the grammar is non-recursive. These grammars produce only one word, which is the uncompressed graph. We denote the uncompressed graph of the SL-HR grammar $H$ by $\psi(H) \in \mathbb{G}$.

\begin{figure*}
    \centering
    \includegraphics[width=\linewidth]{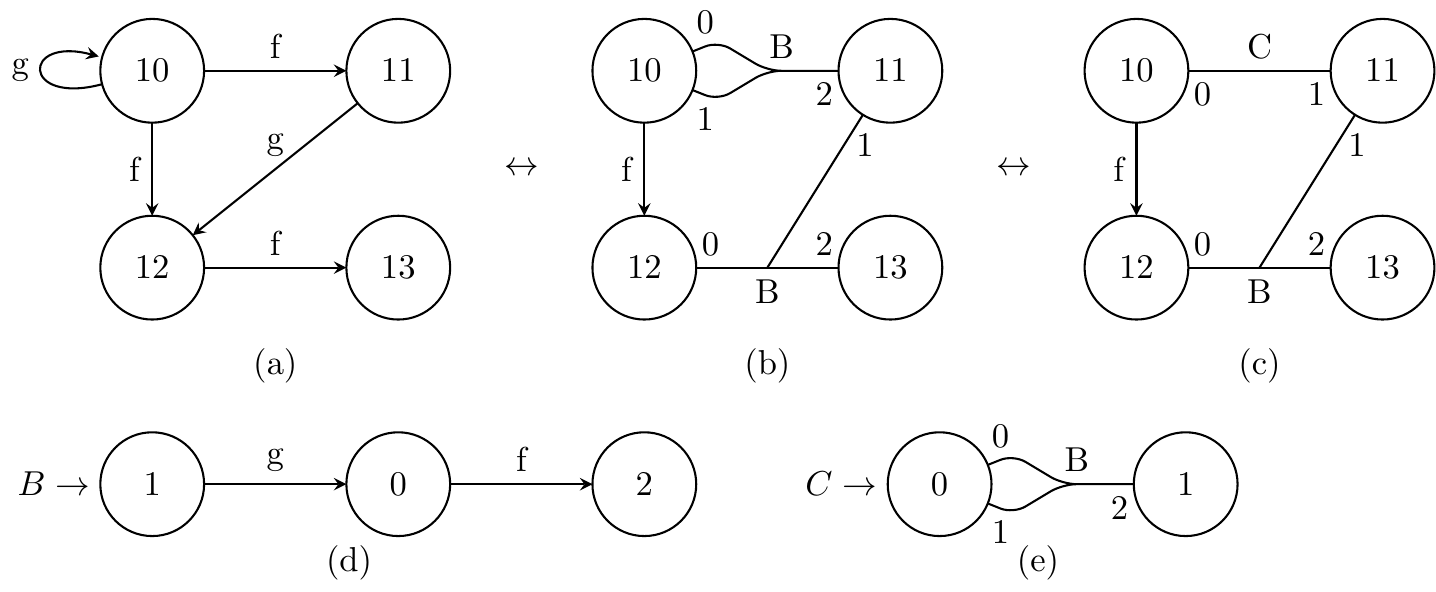}
    \caption{Example hypergraph replacement grammar. From subfigure (a) to (b), we replace all occurrences of the digram $d=((\firstTerminalName,~1), (\secondTerminalName,~0))$ and introduce the rule $\firstNonterminalName$ shown in subfigure (d). In the digram $d$, $(\firstTerminalName,~1)$ means an incoming edge with label $\firstTerminalName$ and $(\secondTerminalName,~0)$ means an outgoing edge with label $\secondTerminalName$. The reverse step from (b) to (a) is called expanding and replacing an occurrence. From (b) to (c), we replace the loop of the edge $\firstNonterminalName(10, 10, 11)$ by introducing the rule $\secondNonterminalName$ shown in (e), which uses the rule $\firstNonterminalName$.  
    }\label{figure:example}
\end{figure*}

However, ITR differs from the approaches of Maneth et al.~\cite{Maneth2016} and Röder et al.~\cite{roeder2021} by our succinct encoding of the edges that replaces loops implicitly. A \textit{loop} is defined as an edge that has more than one connection to the same node. In \autoref{figure:example} (a) and (b) there are loops, both at node $10$. Because loops are replaced in our succinct encoding, our digram definition does not need to distinguish whether or not edges have loops, which is another difference to the approaches of Maneth et al.~\cite{Maneth2016} and Röder et al.~\cite{roeder2021}.

To define digrams, we first define the \textit{incidence-type} $i$ as the pair $(a, m)$ of an edge label $a\in L$ and a connection-type $m \in \mathbb{N}$ with $m < rank(a)$. We define the set of all incidence-types $IT = \{(a, m) \in L \times \mathbb{N} \text{ | }0 \leq m < rank(a)\}$.

A \textit{digram} $d$ is a pair of two possibly equal incidence-types $(a_1, m_1), (a_2, m_2) \in IT$. An \textit{occurrence} of a digram $d = ((a_1, m_1), (a_2, m_2))$ is a pair of two edges $(e_1, e_2)$ with $e_1 \neq e_2$, $label(e_1) = a_1$, $label(e_2) = a_2$, and $e_1[m_1] = e_2[m_2]$. So, the two edges $e_1$ and  $e_2$ are different, fit to the labels of $d$, and share a common node. Let $\mathbb{D}$ be the set of all digrams.

For example, in \autoref{figure:example} (a), for the digram $d = ((\firstTerminalName, 1), (\secondTerminalName, 0))$, we find the occurrences $o_1 = (\firstTerminalName(11, 12), \secondTerminalName(12, 13))$ and $o_2 = (\firstTerminalName(10,  10), \secondTerminalName(10, 11))$. Compression replaces the edges of the occurrences $o_1$ and $o_2$ by $\firstNonterminalName(12, 11, 13)$ and $\firstNonterminalName(10, 10, 11)$, as shown in \autoref{figure:example} (b). In addition, the rule $\firstNonterminalName \rightarrow (\{0, 1, 2\}, \{\firstTerminalName(1, 0), \secondTerminalName(0, 2)\})$, which is visualized in \autoref{figure:example} (d), is added to the grammar. We do not additionally replace $o_3 = (\firstTerminalName(10, 10), \secondTerminalName(10, 12))$, because $o_3$ shares an edge with $o_2$, thereby we cannot replace both, $o_2$ and $o_3$.

The reverse step of replacing a digram occurrence is \textit{expanding} and replacing. Expanding $\firstNonterminalName(12, 11, 13)$ applies the rule $\firstNonterminalName \rightarrow (\{0, 1, 2\}, \{\firstTerminalName(1, 0)\, \secondTerminalName(0, 2)\})$ with its formal node parameters $0, 1, 2$ to the actual nodes $12, 11, 13$ which generates the set $\{\firstTerminalName(11, 12), \secondTerminalName(12, 13)\}$ of edges. By expanding both edges with label $\firstNonterminalName$ in the graph of \autoref{figure:example} (b) and replacing them with the generated sets of edges, we get the decompressed graph of \autoref{figure:example} (a). 


An implication of our definition of digrams is that we only have 3 shapes of digrams for two edges of rank 2 in contrast to 33 or more shapes of Röder et al.~\cite{roeder2021}. Our three shapes have a given common node and either two outgoing edges, two incoming edges, or one incoming and one outgoing edge. In comparison to Röder et al.~\cite{roeder2021}, we save less space by replacing a single occurrence, but we replace more occurrences of a single digram.

\section[Compression]{Compression}

We separate the graph structure from the text and establish the connection between both parts by IDs, which is common practice~\cite{brisaboa2014,fernandez2013,hernandes2015,Maneth2016,roeder2021}. 

The RePair algorithm consists of two steps: \textit{Replace Digrams} (shown in Algorithm~\ref{algo:replace_digrams}) and \textit{Prune} unefficient rules. We omit presenting the Prune step because it is a straight-forward adaptation from String RePair~\cite{larsson2000}. We discuss the steps Count (Line 1), Update Count (Line 6), and Replace (Line 5).

\begin{algorithm}
    \caption{\(replaceDigrams\)}\label{algo:replace_digrams}
    \KwIn{grammar}
    \state{\text{Count occurrences of digrams}}
    \assign{mfd}{\text{most frequent digram}}
    \While{\text{replacing the mfd reduces grammar size}}
    {
        \state{\text{Define a new rule for $mfd$ with a new nonterminal } A_{mfd}}
        \state{\text{Replace each occurrence of $mfd$ with an edge with label } A_{mfd}}
        \state{\text{Update digram count}}
        \assign{mfd}{\text{select next most frequent digram}}
    }
\end{algorithm}

Line 1, Count: To get the maximum number of non-over\-lapping occurrences of a digram, Maneth et al.~\cite{Maneth2016} mention an algorithm with runtime $\bigO(|V|^2|E|)$. Instead, we approximate the count of digrams in two steps. Our digrams consist of two incidence-types, so we first count the frequency of incidence-types at each node by a single scan over all edges. The result is a mapping $c:V \times IT \rightarrow \mathbb{N}$. For example, in \autoref{figure:example} (a), the node $10$ has two outgoing edges with label $\secondTerminalName$, so $c(10, (\secondTerminalName, 0)) = 2$. To define $count:\mathbb{D} \rightarrow \mathbb{N}$, we use $c$ in the following way: For every node $v$ and for every two incidence-types $i_1$ and $i_2$ occurring at the node $v$, we estimate the number of occurrences $count_v(i_1, i_2)$ of the digram $(i_1, i_2)$ at the node $v$ by

\[count_v(i_1, i_2) := \begin{cases}\min(c(v, i_1), c(v, i_2))                     & \text{ if } i_1 \neq i_2 \\
             \left\lfloor \frac{c(v, i_1)}{2} \right\rfloor & \text{ if } i_1 = i_2.\end{cases}\]

In \autoref{figure:example} (a), we have for example $c(10, (\secondTerminalName, 0)) = 2$ and $c(10, (\firstTerminalName, 0)) = 1$. From these two values of $c$, we obtain that at node $10$, there is one occurrence of each of the digrams ($(\secondTerminalName, 0)$,$(\secondTerminalName, 0)$) and ($(\secondTerminalName, 0)$,$(\firstTerminalName, 0)$) and there is no occurrence of the digram $((\firstTerminalName, 0), (\firstTerminalName, 0))$. We define $count:\mathbb{D} \rightarrow \mathbb{N}$ by \[count((i_1, i_2)) = \sum_{v\in V} count_v(i_1, i_2)\] See \hyperref[appendix_a]{Appendix A} for the accuracy of estimation of $count$.


Line 6, Update Count: We store both, $c$ and $count$, from the step Count of Line 1. When we replace an edge $e$ with $label(e)=a$, we consider all nodes $v$ of $e$ and all incidence-types $i_1 = (a, m)$ such that $e[m] = v$ and reduce $c(v, i_1)$ by 1 accordingly. Then, we reduce the number $count((i_1, i_2))$ of all digram occurrences of digrams $d = (i_1, i_2)$ for all incidence-types $i_2$ with $c(v, i_2)>0$ by 1 if and only if
\[i_1 \neq i_2 \text{ and } c(v, i_1) \leq c(v, i_2) \text{ or } i_1 = i_2 \text{ and } c(v, i_1) \text{ is even}.\]
This adjustment has the same result as if we would count the number of digrams calling Line 1 of Algorithm 1 again except for the replaced digram. We do not replace a digram twice, consequently, we remove the current digram from the counting after the update. The steps to increase the counts for the new nonterminal edge are analogous.

Line 5, Replace: We find occurrences of a digram $d = ((a_1, m_1), (a_2, m_2))$ by a left-to-right scan of the edge list and by saving pointers to edges that have one of the labels of $d$. Let $e_1$ have $label(e_1)=a_1$. Then, we store a pointer to $e_1$ according to the node $e_1[m_1]$. If we already had a pointer to $e_2$ with $label(e_2)=a_2$, $e_2[m_2]=e_1[m_1]$, and $e_1\neq e_2$, $(e_1, e_2)$ is an occurrence of $d$. In this case, we delete the pointers to $e_1$ and $e_2$ and replace $e_1$ and $e_2$ with a new hyperedge having label $A_d$. We require $e_1 \neq e_2$ to avoid replacing a loop as an occurrence. In \autoref{figure:example} (a), we replace two occurrences of the digram $((\firstTerminalName, 1), (\secondTerminalName, 0))$, that occur at node $10$ and at node $12$ with one edge with label $\firstNonterminalName$ each, yielding the graph of \autoref{figure:example} (b). 

\subsection*{Succinct Encoding}\label{paragraph:succint_encoding}
\begin{figure*}
    \centering
    \includegraphics[width=\linewidth]{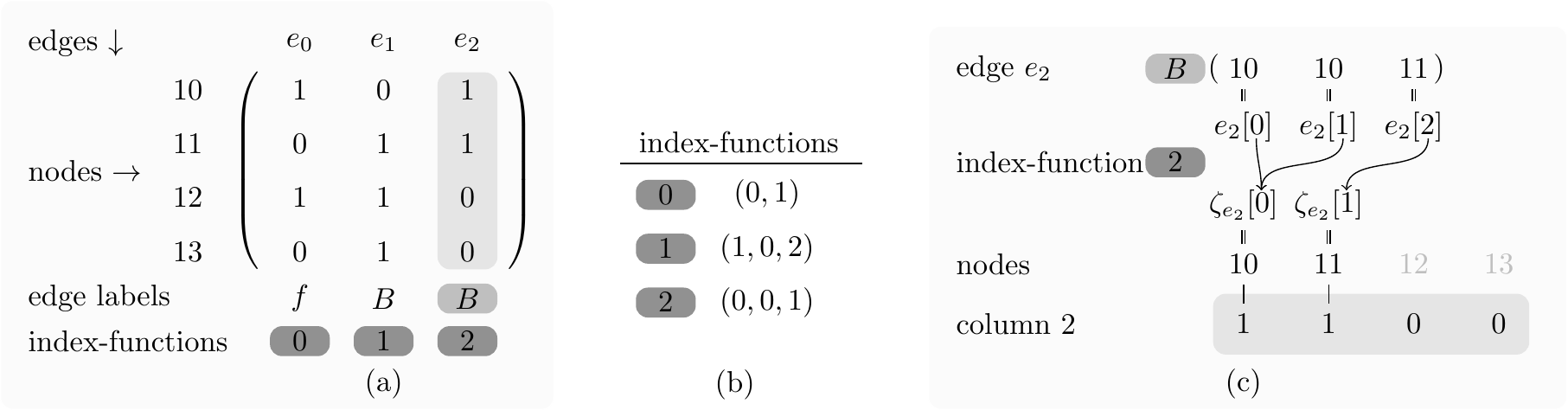}
    \caption{(a) illustrates the implementation of the succinct encoding. One edge in the start rule is represented by three parts: a column of the incidence matrix $M$, an edge label, and an ID of an index-function.\@ (b) shows the IDs and the corresponding index-functions, and (c) shows how the index-function 2 stores the order and the repetitions of nodes $\zeta_{e_2}=[10, 11]$ of edge $e_2$.}\label{figure:succinct_encoding}
\end{figure*}

Our encoding of the start graph $G_S$ is based on $k^2$-trees~\cite{Brisaboa2009} of the incidence-matrix $M$ of $G_S$ and is shown in \autoref{figure:succinct_encoding}. First, we sort the edges by the ID of their label and encode the monotonically increasing list of the IDs by the Elias-Fano-encoding~\cite{vigna2013}. The incidence-matrix $M$ has a $1$ in row $i$ and column $j$ if and only if edge $j$ is connected to node $i$. For example, the edge $e_2 = \firstNonterminalName(10, 10, 11)$ shown in \autoref{figure:succinct_encoding} () is represented by the column $e_2$ of $M$ in \autoref{figure:succinct_encoding} (a). $M$ contains a $1$ at the positions $(10, 2)$ and $(11, 2)$, but it does not contain the information how often and at which connection-types the node $10$ occurs in edge $e_2$. We introduce the \textit{index-function} to close this information gap.

Let $\zeta_e$ be the duplicate-free and sorted list of nodes of $e$. In the example of \autoref{figure:succinct_encoding} (c), $\zeta_{e_2}$ is $[10, 11]$. Let $k$ be the length of $\zeta_e$. The \textit{index-function} $\pi_e$ 
maps each connection-type $m$ of the edge $e$ to its index in the list $\zeta_e$. Formally, $e[m] = \zeta_e[\pi_e(m)]$. We write $\pi_e$ as the tuple ${(\pi_e(0), \ldots, \pi_e(rank(e)-1))}$. In the example of \autoref{figure:succinct_encoding} (b), $\pi_{e_2}=(0, 0, 1)$. Thereby, each edge is uniquely reconstructable by its column in $M$ together with its index-function and its label. Instead of saving the same index-function multiple times, we assign IDs to all index-functions. We encode $\pi_e$ as the sequence $\delta(rank(e)-1)$$\delta(\pi(0))\ldots\delta(\pi(rank (e)-1))$ using the $\delta$-code~\cite{elias1975}.

Like Maneth et al.~\cite{Maneth2016}, we encode all graphs of rules except the start graph by only encoding the right-hand sides of the rules because the order of rules determines their nonterminal. For each right-hand side, we first encode the number of edges and then, for each edge, the label and its nodes. For example, the sequence of edges $d (0, 1)$ and $\firstNonterminalName{} (1, 2, 0)$ is encoded by $\delta{} (2)$ $\delta{} (d)\delta{} (0)\delta{} (1)$ $\delta{} (\firstNonterminalName)\delta{} (1)\delta{} (2)\delta{} (0)$.

\subsection*{Handling Loops in the Grammar}\label{paragraph:unigrams}
In \autoref{figure:example} (b), we have an edge $e = \firstNonterminalName(10, 10, 11)$ that is connected twice to node $10$ and thereby is a loop. We could introduce an extra rule $\secondNonterminalName{} \rightarrow{} (\{0, 1\}, \{\firstNonterminalName(0, 0, 1)\})$ as in \autoref{figure:example} (e) and replace the edge $e$ by $ \secondNonterminalName(10, 11)$ yielding the graph of \autoref{figure:example} (c). This is a tradeoff: We introduce more rules, but we reduce the number of nodes that are used as a parameter in rules by 1 for each occurrence of such a loop. An evaluation of the implementation shows that these extra rules do not improve compression, because the index-function that is used in the succinct encoding also removes the duplicate parameters. Therefore, we do not replace loops by introducing extra rules, and also because skipping this step reduces the time needed to compress a graph.

\subsection*{Answering triple queries}


Simple graph queries written as triples are more general than neighborhood queries. A triple query asks for all edges with label P from node S to node O, written as \texttt{S P O}. Each component \texttt{S}, \texttt{P}, or \texttt{O} is either bound or unbound. For example, the query \texttt{10 f ?} asks for all edges with label f from node 10 to any node. 

Given an SL-HR grammar $H$ with start rule $S \rightarrow{} G_S$, we answer a triple query based on its bound parameters. We use a set $Z$ to store intermediate results.
\begin{itemize}
    \item Case 1, \texttt{S} or \texttt{O} are bound: Let $r$ be the value of \texttt{S}, or of \texttt{O} if \texttt{S} is unbound. To compute set $Z$, we first decompress the row $r$ of the incidence-matrix $M$ of $G_S$. This is done without decompressing the whole incidence-matrix~\cite{Brisaboa2009}. Each column $j$ with a 1 in row $r$ represents an edge $e_j$ that is connected to node $r$, and we add the edge $e_j$ to $Z$.
    \item Case 2, only \texttt{P} is bound: An edge with label \texttt{P} can occur in the start graph or can be generated by a nonterminal edge $e$ with label $A$. We lookup in a matrix $NT$, whether or not $e$ directly or indirectly creates edges with label \texttt{P}. The matrix $NT$ contains a row for each nonterminal label $A_y$ in the grammar and a column for each terminal label $a_x$, and there is a 1 in $NT$ if and only if $A_y$ generates at least one edge with $a_x$ as label. $NT$ is compressed by $k^2$-trees. If $e$ generates edges with label \texttt{P}, we add $e$ to $Z$. We use the binary search on the sorted list of edge labels of the start graph.
    \item Case 3, the query is \texttt{?S\@ ?P\@ ?O}: The query is equivalent to the decompression. We add all edges of the start rule to $Z$.
\end{itemize}
In all three cases, while there is an edge $e \in{} Z$: 

\begin{itemize}
\item If $e$ is a nonterminal and
\begin{itemize}
    \item[---] (\texttt{S} is unbound or $e$ is connected to \texttt{S} for any connection-type) and
    \item[---] (\texttt{O} is unbound or $e$ is connected to \texttt{O} for any connection-type) and
    \item[---] (\texttt{P} is unbound or $NT$ has the value $1$ at row $label (e)$ and column \texttt{P}),
    \item[$\rightarrow$] expand $e$ and extend $Z$ by the set of edges generated by expanding $e$.
\end{itemize}
\item If $e$ is a terminal and $e$ is a solution to the query, output $e$.
\item In all cases, remove $e$ from $Z$, and proceed with a new $e \in{} Z$.
\end{itemize}

\subsection*{ITR+: Improved Compression of Frequent Node Labels}

\textit{ITR+} is an extension of ITR that increases the compression for graphs that contain different nodes with the exact same label. For example, the ttt-win graph of the game tic-tac-toe contains only 3 node labels, x and o for the players, and b for a blank field. Like gRePair~\cite{Maneth2016} and RDFRePair~\cite{roeder2021}, ITR stores $|V| = 5634$ representations of these three labels in the dictionary, because the RDF representations are slightly different for all $5634$ nodes. However, ITR+ stores only three dictionary entries for the labels x, o and b. ITR+ inserts edges of rank 1 into the graph; for example, x(1) means that the node with ID 1 has the label x. Thereby, ITR+ stores only three labels for these edges in the dictionary instead of storing $5634$ labels for nodes. Furthermore, edges of rank 1 can be part of a digram replacement with other edges: ITR+ compresses frequent subgraphs of node labels and edge labels into single nonterminals. Thus, ITR+ reduces the compression ratio of the ttt-win graph to 14.23\% of the uncompressed graph, which outperforms all other approaches. On the other hand, ITR+ increases the number of edges from $|E|$ to $|E| + |V|$.

The gRePair implementation is not capable of using edges of rank 1 as terminals and RDFRePair is designed only for edges of rank 2. Therefore, the ability to treat frequent node labels as hyperedges of rank 1 and thereby compress the graph better is unique to ITR+.

\section[Experimental Results]{Experimental Results}

\begin{table*}
    \footnotesize
    \begin{minipage}{0.48\linewidth}
        \begin{tabular}{llll}
            Name       & Lang. & Query Type   & Paper 
            \\
            \hline
            ITR        & C        & triple      & $-$ 
            \\
            RDFRePair  & Java     & -            & \cite{roeder2021} 
            \\
            gRePair    & Scala    & Neighborhood & \cite{Maneth2016} 
            \\
            HDT-java   & Java     & SPARQL       & \cite{fernandez2013} 
            \\
            HDT++      & C++      & triple       & \cite{hernandes2015} 
            \\
            $k^2$-java & Java     & SPARQL       & \cite{Brisaboa2009} 
            \\
            $k^2$++    & C++      & -            & \cite{Brisaboa2009} 
            \\
        \end{tabular}
        \vfill
        \begin{center}
            (a)
        \end{center}
    \end{minipage}
    \begin{minipage}{0.42\linewidth}
        \begin{tabular}{lrrr}
            File 
                & $|V|$    & $|E|$    
                & $|T|$               \\ \hline
            homepages-en 
                & 98665    & 50000    
                & 1                   \\
            geo-coordinates-en 
                & 46107    & 50000    
                & 4                   \\
            jamendo 
                & 396531   & 1047951  
                & 25                  \\
            wikidata 
                & 10051660 & 42922799 
                & 635                 \\
            archiveshub 
                & 280556   & 1361816  
                & 139                 \\
            scholarydata-dump 
                & 140042   & 1159985  
                & 84                  \\
            \hline
            chess-legal 
                & 76272    & 113039   
                & 12                  \\
            ttt-win 
                & 5634     & 10016    
                & 3                   \\
            \hline
            WikiTalk 
                & 2394385  & 5021410  
                & 1                   \\
            NotreDame 
                & 325729   & 1497134  
                & 1                   \\
            CA-AstroPh 
                & 18772    & 396160   
                & 1                   \\
        \end{tabular}
        \begin{center}
            (b)
        \end{center}
    \end{minipage}
    \caption{(a) Compared implementations of the compression algorithms. (b) Datasets used for testing, sorted by the edge per node ratio $\frac{|E|}{|V|}$. The number of labels of a dataset is listed in column $|T|$.}\label{table:datasets}
\end{table*}

We investigate the time needed to answer a neighborhood query or a triple query. We compare the results of the compressors listed in \autoref{table:datasets} (a). We test all implementations on a Debian 5.10.209-2 machine with 128GB RAM and 32 Cores Intel(R) Xeon(R) Platinum 8462Y+ @ 2.80GHz. For the tests, we use the RDF graphs homepages-en, geo-coorindates-en, jamendo, wikidata, archiveshub, scholarydata-dump as in Röder et al.~\cite{roeder2021}, the version graphs ttt-win and chess-legal from SUBDUE\footnote{\url{https://ailab.wsu.edu/subdue/download.htm}}, and the webgraphs CA-AstroPh, NotreDame, and WikiTalk from the Standford Large Network Dataset Collection~\cite{snapnets}. 

ITR is implemented\footnote{The implementation of ITR is available at \url{https://github.com/adlerenno/IncidenceTypeRePair}} in C. For the $k^2$-tree approach, we use the implementations of Röder et al.~\cite{roeder2021} as in their paper in Java and C++, $k^2$-java and $k^2$++, respectively. We include two implementations of HDT, namely HDT-java~\cite{fernandez2013} and HDT++~\cite{hernandes2015}. We omit RDF engines because RDF engines mostly cannot handle multiple nodes with the same label and thus, RDF is only a subclass of graphs we look at.

\begin{figure*}
    \centering
    \includegraphics[width=0.92\linewidth]{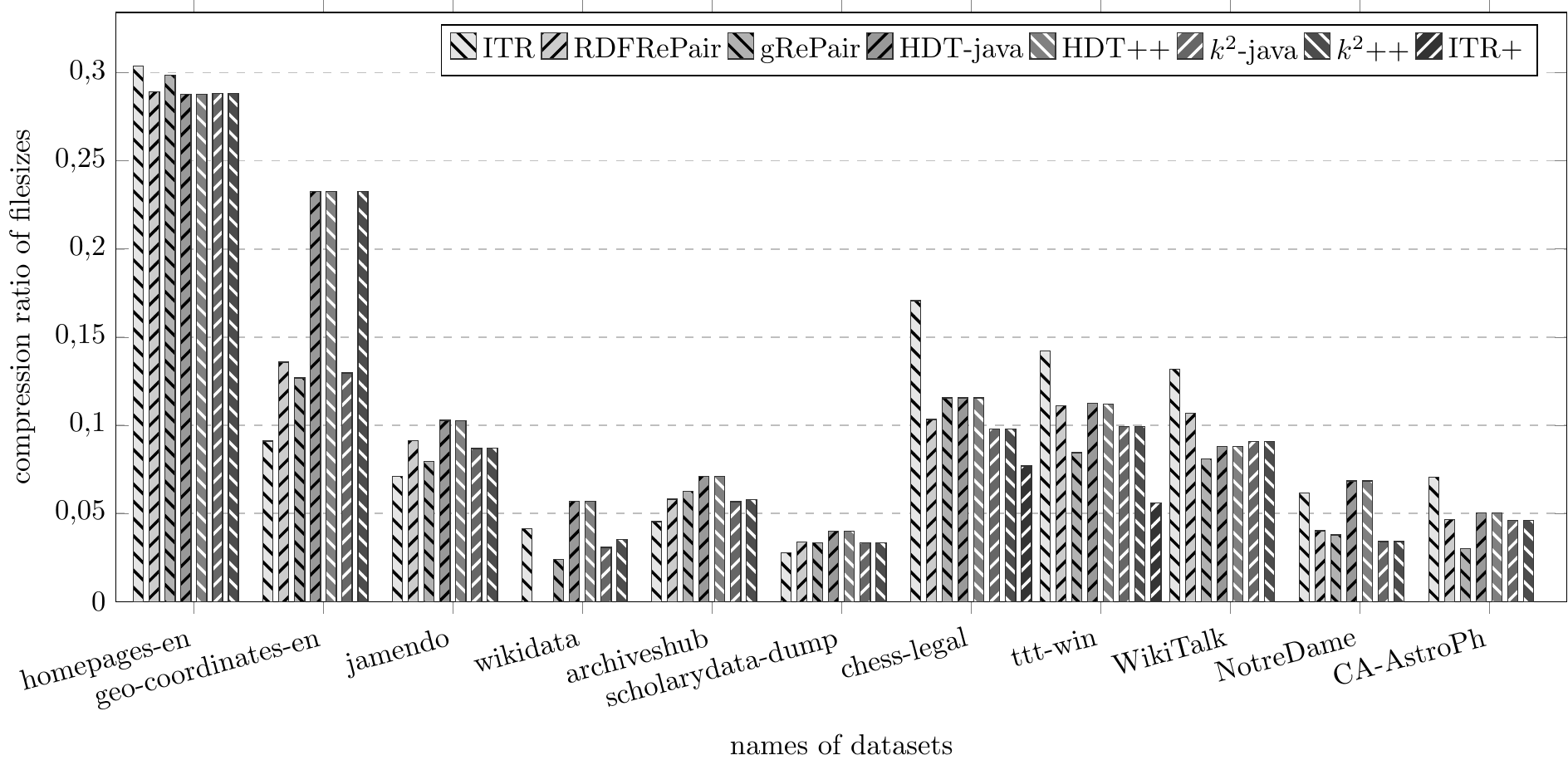}

    \caption{The compression ratio is the file size of the compressed graph divided by the file size of the uncompressed input file. We stopped RDFRePair on wikidata after 6 days. For ITR+, only chess-legal and ttt-win use the same node label for multiple nodes.}\label{figure:compression_ratio}
\end{figure*}

In \autoref{figure:compression_ratio}, we see that ITR compresses better than the existing solutions on some datasets like geo-coordinates, and archiveshub, but other approaches compress better on the network and version graphs due to a non-optimal dictionary compression.

\begin{figure*}[!tb]
    \centering



    \includegraphics[width=0.92\linewidth]{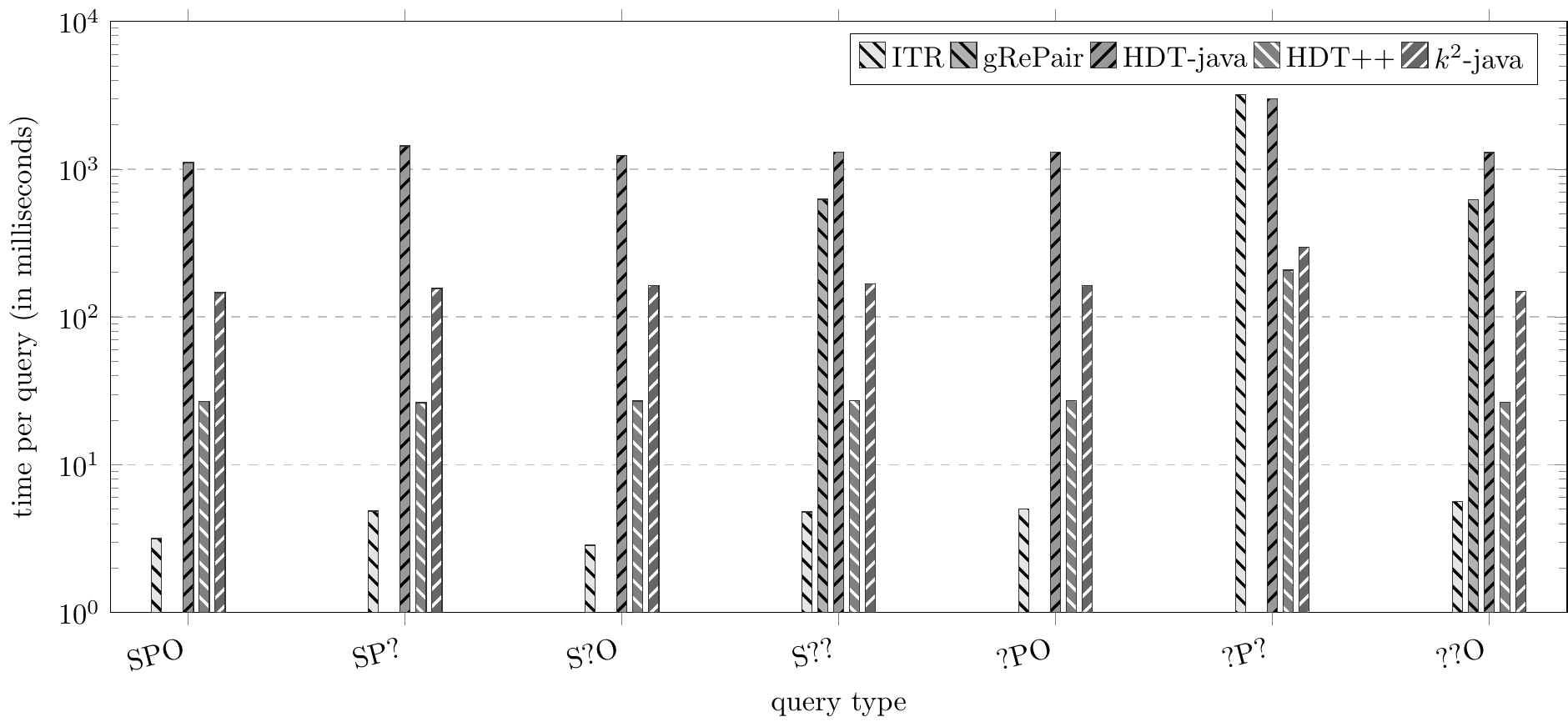}
    \caption{Average runtime in milliseconds of 500 queries each on the jamendo graph. Implementations that do not appear in the legend do not support any type of query. gRePair can only answer neighborhood queries (\texttt{S ?\@ ?} and \texttt{?\@ ?\@ O}).
    }\label{figure:neighborhood_query_time}
\end{figure*}

For the runtime comparison summarized in \autoref{figure:neighborhood_query_time}, we include all times needed for IO-operations as not all compressors output their time for performing the query without IO-operations. 
We exclude the build time and disk space used by HDT-java and HDT++ for additional indices.

Due to the different supported query types, we compare the runtime by using queries that are equal. As the gRePair implementation does not support queries with edge labels, gRePair is only included in the comparisons of \texttt{S ?\@ ?} and \texttt{?\@ ?\@ O}. We use \texttt{SELECT ?pr ?ob WHERE \{ v ?pr ?ob \}} as SPARQL query and \texttt{v ?P ?O} as triple query for an outgoing neighborhood query of the node $v$. We perform 500 queries of each query type on each file and show the average time needed to answer the queries.

In \autoref{figure:neighborhood_query_time}, we see that gRePair, HDT-java, and $k^2$-java as Java or Scala implementations are outperformed by the C and C++ implementations. ITR is about 2 to 6 times faster than HDT++ except for \texttt{?\@ P ?} queries. Furthermore, ITR is more than 100 times faster than gRePair. In \autoref{figure:neighborhood_query_time}, we presented the results for the jamendo graph, but the time ratios of ITR to other approaches are similar for all tested graphs.



\section[Conclusion]{Summary and Conclusion}

We presented the ITR graph compressor based on a new type of digrams and on index-functions. Furthermore, ITR+ can use hyperedges of rank 1 for frequent node labels to reduce the needed dictionary size to store node labels. Our evaluation of the runtime of triple queries shows that ITR performs all queries except \texttt{?\@ P ?} significantly faster than the existing variants of RePair on graphs, namely gRePair and RDF\-RePair, and faster than $k^2$-tree and HDT provided on different implementations.  

\subsection*{Acknowledgements}

We would like to thank Fabian Röthlinger for his support in implementing ITR.

\appendix
\section*{Appendix A: On the Accuracy of the Count Estimate} \label{appendix_a}

In Appendix A, we investigate the accuracy of the presented counting of digrams. Let $G$ be a fixed graph and $d = ((a, k), (b, l))$ be a fixed digram. Let $O$ be a set of non-overlapping occurrences of $d$ and $O$ be maximal in size. We show that $count(d) \geq |O|$, and $count(d) = |O|$ if either $a \neq b$ or $k = l$.

We define $O_v = \{(e_1, e_2) \in O | e_1[k] = e_2[l] = v\}$ as the subset of $O$ that has the node $v$ as shared node defined by the incidence-types of digram $d$. For two nodes different nodes $v_1$ and $v_2$, $O_{v_1} \cap{} O_{v_2} = \emptyset$, because we get $v_1 = e_1[k] = v_2$ for all $e_1 \in O_{v_1} \cap{} O_{v_2}$ as an obvious contradiction otherwise. Hence each occurrence $(e_1, e_2) \in O$ is in $O_{e_1[k]}$ ($=O_{e_2[l]}$), the sets $O_v$ for all nodes $v \in V$ form a disjoint union of $O$, so $O = \bigcupdot_{v \in V} O_v$.

Regarding a single fixed node $v$, $E_1 = \{e \in E | label(e) = a \text{ and } e[k] = v\}$ and $E_2 = \{e \in E | label(e) = b \text{ and } e[l] = v\}$ contain possible edges for the first and second incidence-type of the digram $d$, respectively. If the two incidence-types are equal, $E_1 = E_2$ and $\left\lfloor \frac{|E_1|}{2} \right\rfloor$ is the maximum number of pairs that can be drawn from $E_1$ without repetition. If the incidence-types are different, $E_1 \neq E_2$ and consequently $min(|E_1|, |E_2|)$ is the maximum number of non-overlapping pairs of edges. We define $count_v$ accordingly.

\begin{align*}
    \left| O\right| &= \left|\bigcupdot_{v \in V} O_v \right| \\
    &= \sum_{v \in V} |O_v| \\
    &\leq \sum_{v \in V} count_v(d) \\
    &= count(d)
\end{align*}

This shows that our estimate of the number of digram occurrences is an upper bound. Assuming $a \neq b$ or $k = l$, each edge $e \in E$ with label $a$ (or $b$) belongs to exactly one node $e[k]$ (or $e[l]$). Only at $e[k]$ (or $e[l]$) it is possible to form a digram using the edge $e$. $|O_v| = count_v(d)$ follows. 

Only the case $a = b$ and $k \neq l$ violates in two ways our counting: Overlapping occurrences with different shared nodes or loops with the incidence-types of the digram are not counted correctly. An completely accurate count would take $\bigO(|E|^2)$~\cite{Maneth2016}, which is infeasible.

%
%
%
\section*{References}
\bibliographystyle{IEEEbib}
\bibliography{biblio}

\begin{thebibliography}{10}

\bibitem{larsson2000}
N.~J. Larsson and A.~Moffat,
\newblock ``{Off-line dictionary-based compression},''
\newblock {\em Proceedings of the IEEE}, vol. 88, no. 11, pp. 1722--1732, 2000.

\bibitem{gasieniec2005}
L.~Gasieniec, R.~M. Kolpakov, I.~Potapov, and P.~Sant,
\newblock ``{Real-Time Traversal in Grammar-Based Compressed Files.},''
\newblock in {\em DCC}. Citeseer, 2005, p. 458.

\bibitem{Lohrey2011}
M.~{Lohrey}, S.~{Maneth}, and R.~{Mennicke},
\newblock ``{Tree Structure Compression with RePair},''
\newblock in {\em 2011 Data Compression Conference}, 2011, pp. 353--362.

\bibitem{lohrey2016}
M.~Lohrey, S.~Maneth, and C.~P. Reh,
\newblock ``{Traversing Grammar-Compressed Trees with Constant Delay},''
\newblock in {\em 2016 Data Compression Conference (DCC)}, 2016, pp. 546--555.

\bibitem{Maneth2016}
S.~Maneth and F.~Peternek,
\newblock ``{Compressing graphs by grammars},''
\newblock in {\em 2016 IEEE 32nd International Conference on Data Engineering (ICDE)}. IEEE, 2016, pp. 109--120.

\bibitem{roeder2021}
M.~R{\"o}der, P.~Frerk, F.~Conrads, and A.-C. Ngonga~Ngomo,
\newblock ``{Applying Grammar-Based Compression to RDF},''
\newblock in {\em The Semantic Web}, Cham, 2021, pp. 93--108, Springer International Publishing.

\bibitem{fernandez2013}
J.~Fern{\'a}ndez, M.~A. Mart{\'i}nez-Prieto, C.~Gutierrez, A.~Polleres, and M.~Arias,
\newblock ``{Binary RDF Representation for Publication and Exchange (HDT)},''
\newblock {\em Journal of Web Semantics}, vol. 19, pp. 22--41, 03 2013.

\bibitem{Brisaboa2009}
N.~R. Brisaboa, S.~Ladra, and G.~Navarro,
\newblock ``{k2-Trees for Compact Web Graph Representation},''
\newblock in {\em String Processing and Information Retrieval}, Berlin, Heidelberg, 2009, Springer Berlin Heidelberg.

\bibitem{brisaboa2014}
N.~R. Brisaboa, S.~Ladra, and G.~Navarro,
\newblock ``{Compact representation of Web graphs with extended functionality},''
\newblock {\em Information Systems}, vol. 39, pp. 152--174, 2014.

\bibitem{hernandes2015}
A.~Hern{\'a}ndez-Illera, M.~A. Mart{\'i}nez-Prieto, and J.~D. Fern{\'a}ndez,
\newblock ``{Serializing RDF in Compressed Space},''
\newblock in {\em 2015 Data Compression Conference}, 2015, pp. 363--372.

\bibitem{vigna2013}
S.~Vigna,
\newblock ``{Quasi-Succinct Indices},''
\newblock in {\em Proceedings of the Sixth ACM International Conference on Web Search and Data Mining}, New York, NY, USA, 2013, WSDM '13, p. 83–92, Association for Computing Machinery.

\bibitem{elias1975}
P.~Elias,
\newblock ``{Universal codeword sets and representations of the integers},''
\newblock {\em IEEE Transactions on Information Theory}, vol. 21, no. 2, pp. 194--203, 1975.

\bibitem{snapnets}
J.~Leskovec and A.~Krevl,
\newblock ``{SNAP Datasets}: {Stanford} large network dataset collection,'' \url{http://snap.stanford.edu/data}, June 2014.

\end{thebibliography}

\end{document}